\documentclass[
aps,%
11pt,%
final,%
notitlepage,%
oneside,%
twocolumn,%
nobibnotes,%
nofootinbib,%
superscriptaddress,%
noshowpacs,%
centertags]%
{revtex4}
\usepackage{caption2}
\usepackage{graphicx}
\usepackage{epsfig}

\begin{document}

\title{The stellar-wind envelope around
  the supernova XRF/GRB\,060218/SN\,2006aj massive progenitor star}

\author{\firstname{E.}~\surname{Sonbas}}
\affiliation{Special Astrophysical Observatory of R.A.S., Karachai-Cherkessia,
Nihnij Arkhyz, 369167 Russia}
\affiliation{University of Cukurova, Department of Physics,
 01330 Adana, Turkey }

\author{\firstname{A.~S.}~\surname{Moskvitin}}
\affiliation{Special Astrophysical Observatory of R.A.S., Karachai-Cherkessia,
Nihnij Arkhyz, 369167 Russia}

\author{\firstname{T.~A.}~\surname{Fatkhullin}}
\affiliation{Special Astrophysical Observatory of R.A.S., Karachai-Cherkessia,
Nihnij Arkhyz, 369167 Russia}

\author{\firstname{V.~V.}~\surname{Sokolov}}
\affiliation{Special Astrophysical Observatory of R.A.S., Karachai-Cherkessia,
Nihnij Arkhyz, 369167 Russia}

\author{\firstname{A.}~\surname{Castro-Tirado}}
\affiliation{Instituto de Astrofsica de Andalucia (IAA-CSIC), P.O.
Box 03004, 18080 Granada, Spain}

\author{\firstname{A.}~\surname{de Ugarte Postigo}}
\affiliation{Instituto de Astrofsica de Andalucia (IAA-CSIC), P.O.
Box 03004, 18080 Granada, Spain}
\affiliation{European Southern Observatory (ESO), Chile}

\author{\firstname{J.}~\surname{Gorosabel}}
\affiliation{Instituto de Astrofsica de Andalucia (IAA-CSIC), P.O.
Box 03004, 18080 Granada, Spain}

\author{\firstname{S.}~\surname{Guziy}}
\affiliation{Instituto de Astrofsica de Andalucia (IAA-CSIC), P.O.
Box 03004, 18080 Granada, Spain}

\author{\firstname{M.}~\surname{Jelinek}}
\affiliation{Instituto de Astrofsica de Andalucia (IAA-CSIC), P.O.
Box 03004, 18080 Granada, Spain}

\author{\firstname{T.~N.}~\surname{Sokolova}}
\affiliation{Special Astrophysical Observatory of R.A.S., Karachai-Cherkessia,
Nihnij Arkhyz, 369167 Russia}

\author{\firstname{V.~N.}~\surname{Chernenkov}}
\affiliation{Special Astrophysical Observatory of R.A.S., Karachai-Cherkessia,
Nihnij Arkhyz, 369167 Russia}
\def\g{$\gamma$}

\begin{abstract}
{\bf Abstract.}
In BTA spectra of the supernova SN\,2006aj, identified with the X-ray flash
(XRF) and gamma-ray burst XRF/GRB\,060218/SN\,2006aj, we detected details 
interpreted as hydrogen lines, which is a sign of stellar-wind envelope 
around a massive progenitor star of the gamma-ray burst.
Results of modeling two early spectra obtained with the BTA
in 2.55 and 3.55 days after the explosion of Type Ic supernova
SN\,2006aj (z=0.0331) are presented.
The spectra were modeled in the Sobolev approximation with the
SYNOW code (Branch et al. 2001; Elmhamdi et al. 2006).
In the spectra of the optical afterglow of the X-ray flash XRF/GRB\,060218
we detected spectral features interpreted as
(1) the $H\alpha$ PCyg profile for the velocity  33000\,km\,s$^{-1}$ ---\
a wide and almost unnoticeable deformation of continuum in the range
of $\simeq5600 - 6600\AA$ for the first epoch (2.55 days) and
(2) a part (``remnant'') of the $H\alpha$ PCyg profile in absorption
blueshifted by 24000\,km\,s$^{-1}$ ---\ a wide spectral feature with a 
minimum at $\simeq6100\AA$  (the rest wavelength) for the second
epoch (3.55 days). 
Taking into consideration early BTA observations and spectra obtained with 
other telescopes (ESO Lick, ESO VLT, NOT) before 2006 Feb. 23 UT, it can be 
said that we observe evolution of optical spectra of Type Ic
core-collapse supernova SN\,2006aj during {\it transition}
from the short phase related to the shock breakout to outer layers
of the stellar-wind envelope to spectra of the phase of increasing brightness
corresponding to radioactive heating.
Signs of hydrogen in spectra of the gamma-ray burst afterglow were detected
for the first time.
\end{abstract}


   \maketitle

\section{Introduction}

   On Feb. 18.149, 2006 UT the space observatory \textit{Swift}
detected a peculiar gamma-ray burst (GRB)
with a powerful component of supernova (SN) emission in
spectra and in the light curve of the GRB afterglow. 
Therefore, this burst is simultaneously classified both as GRB\,060218,
and as SN2006aj: GRB\,060218/SN\,2006aj. 
But since in this case the X-ray emission was prevailing in the GRB 
spectrum, the GRB is also classified as XRF (X-Ray Flash),
and the event is denoted either as XRF/GRB\,060218/SN\,2006aj,
or as XRF\,060218/SN\,2006aj.
Below we will mostly use the latter notation to emphasize the circumstance
which is important for this work that in the case of SN\,2006aj
the SN event itself started with observation of a powerful X-ray flash (XRF).

   For this XRF/GRB,
direct observational signs of an early
phase of expansion of shock arising due to explosion of a compact core
of a pre-SN and the shock breakout to the outer boundary of the stellar-wind
envelope surrounding the SN progenitor star
  were obtained for the first time.
During the first 2 hours, the shockwave was observed as the mostly
X-ray flash XRF\,060218, and afterwards,
 when the envelope became optically thin,
 as a powerful ultraviolet (UV) burst with the maximum brightness
 in 11 hours after GRB detection (Campana et. al., 2006; Blustin 2007).
During first 2800 seconds, beside non-thermal emission typical of GRB
afterglows, XRF\,060218 also demonstrates in its X-ray spectrum a powerful
{\it thermal} component whose temperature falls, and with time the emission 
maximum shifts to the UV and optical spectral range (see Fig.1 in
Campana et. al., 2006 and Fig. 2 in Blustin 2007).

   XRF/GRB\,060218 is one of the nearest GRBs with its redshift $z$=0.0331.
In this respect it can be compared to GRB\,030329/SN\,2003dh having the  
redshift $z$=0.1683 which was also identified with the Type Ic SN.
Both events have aroused considerable interest because such coincidences 
(GRB and SN) happen rarely, only once every two-three years (Chapman et
al. 2007), when GRBs revealed SNe starting from the moment of star
core explosion.
It can be said that the study of GRBs is a new phase of investigation
of the same core-collapse SNe, but {\it from the very beginning} of
this remarkable phenomenon.
That is why in these cases early spectral observations turn out to be very
important for understanding the mechanism of both core-collapse SN
explosion itself and GRB source.
When such early observations are successful, the 6-meter telescope,
with its large accumulating area and eastern location relatively other large
European observatories, can play an important role in implementation
of the international program of spectroscopic observations (monitoring)
of quickly-fading optical GRB afterglows.

   Fig. 1 represents the spectra obtained with the 6-meter telescope
(Fatkhullin et al. 2006).
Like the case of GRB\,030329/SN2003dh (Sokolov et al. 2003),
the spectra of XRF\,060218/SN2006aj are among the earliest spectra
obtained with a high signal/noise ratio (as is shown in Table 1),
which permits applying the interpretation methods usually used
for spectra of Type Ic SNe
(Branch et al. 2001; Elmhamdi et al. 2006).

   As was shown by Campana et al. (2006),
both the XRF itself and the UV burst in about 10-11 hours after
XRF\,060218, and then the UV excess in early spectra of the afterglow
can be explained by interaction between the SN shock and a stellar-wind
envelope around the massive progenitor star.
This is a so-called {\it ``shock breakout''} effect
(the shock breaks out through an envelope around a collapsing
and exploding star core).

   Such an effect for core-collapse SNe\,Ib/c and SNe\,II
has been known for a long time already (Colgate, 1968;
Imshennik and Nadezhin, 1989; Calzavara and Matzner, 2004).
It can be observed as a relatively short phase of the SN explosion which,
after beginning with an X-ray flash (XRF), ends with a bright UV burst, 
which announces arrival of the shock to surface of the exploding star.
It can be also said about the shock breakout to upper layers or
to an optically-thin ``surface'' of the extended stellar-wind envelope
surrounding the collapsing and exploding cores of Type Ib or Ic SNe.
Below, the term ``shock breakout'' will be used.
Like two famous SN\,1993J and SN\,1987A, in the case of XRF\,060218/SN2006aj
this effect was observed as a sharp and short early peak in the optical
light curve --- see Fig.2 in Campana et al. 2006 and Fig.2 in this paper.
We mentioned the same effect when explaining the very first spectra of the
GRB\,030329/SN2003dh afterglow obtained in $\simeq $10 hours
after GRB\,030329 (Sokolov 2003).
Due to detecting XRF\,060218/SN2006aj in the beginning of the SN burst,
i.e. before the shock breakout, we managed to observe motion of the shock
inside the stellar-wind envelope (the shock breakout effect) during the 
first two hours as an X-ray flash (XRF) with thermal spectrum.

   From Table 1 it is seen that BTA spectra refer to minimum
in the light curve, i.e. we observe the transition
   from phase 1 of thermal emission related to the shock breakout
to ``surface'' of the stellar-wind envelope (the first peak in Fig.2)
   to phase 2 of the subsequent increasing of SN brightness
with maximum in $\simeq$10 days in Fig.2, which corresponds to
the (non-thermal) radioactive heating of the expanding SN envelope
caused by the decay $^{56}$Ni$\to^{56}$Co$\to^{56}$Fe.

 Thus, during our observations, physical conditions have been changing
quickly, and this paper is dedicated to interpretation of spectra
of this remarkable transitional phase.
 But before comparing observational and theoretical spectra
in Section 2 of this paper, we adduce estimates of typical energy,
temperature, size and velocity which directly follow from results
of \textit{Swift} observations of XRF\,060218 in the X-ray and UV ranges
obtained before the beginning of the BTA spectral observations.

1) {\it Energy of the X-ray and UV flashes:}

   Light curves of the XRT (0.3-10 keV) and UVOT ranges are given
in the paper by Campana et al. (2006).
 Ibidem, there is an estimate of energy radiated in gamma- and X-rays
during the first two hours when the shock has been still moving
(breaking out) through the wind.
 Energy release in the X-ray (XRT) flash
is equal to $\sim 6 \times 10^{49}$ erg.
 Then, in about 8-11 hours, the UV (UVOT) light curve shows a powerful peak
(the first maximum in our Fig.2) caused by {\it the shock breakout}
to the surface or, more exactly, to the outer boundary
of the stellar-wind envelope when it becomes sufficiently transparent.
 It is this UV burst that is the Colgate shock breakout effect
(Colgate 1968).
 Energy released in this UV flash in the Swift/UVOT range
during $\approx$ 28 hours and estimated from data
of the paper by Campana et al. (2006)
gives the number of the same order $\sim 3 \times 10^{49}$ erg,
which directly says about identical nature of the X-ray (XRT)
and UV (UVOT) flashes/bursts.

2) {\it Evolution of temperature: }

   At first, when the shock breaks out through the wind envelope
and energy is released mostly in X-rays, the thermal spectrum
corresponds well to the temperature
kT $\approx$ 0.17 keV ($\simeq$2x$10^6$\,K).
 The temperature falls to 0.03 - 0.05 keV (350000 - 580000\,K) or lower,
by the time of the UV (UVOT) flash in $\approx$ 11 hours after the GRB
(the first maximum in our Fig.2)
taking into consideration the kT measurement uncertainties
with Swift/UVOT in this period
(only a part of the black-body peak gets into the UVOT range,
see Blustin, 2007).
 Then the temperature of the thermal component decreased down to
$\simeq$43000\,K or to kT $\approx$ 3.7 +1.9/-0.9 eV
(Campana et al. 2006, Fig. 3) in the subsequent $\approx$22.5 hours.
 Thus, it is thought that
the temperature can be even lower, less than 10000\,K,
by the moment of our first spectral observation
(2.55 days after GRB, see Table 1),
in accordance with estimates
of the temperature decrease rate adduced in the paper
by Campana et al..

3) {\it Size of the stellar-wind envelope: }

   From the data given in the paper by Campana et al. (2006),
one can also estimate radius of the wind envelope surrounding 
the Wolf-Rayet (WR) star before the explosion.
 (For example, the arguments in favor of the WR star surrounded by an extended
stellar-wind envelope as a progenitor star
of XRF\,060218/SN\,2006aj are given by Blustin, 2007).
 It is natural to connect the size of this envelope to the bright UV (UVOT)
flash observed in $\simeq $11 hour after GRB.
 At this moment the shock which was observed before that only in X-rays
becomes visible in UV and optical at last, since layers of the wind envelope
{\it over} the shock become optically thin and the shock breaks out
to the ``surface'' (more exactly, to upper layers) of this extended envelope
related to the GRB progenitor star or pre-SN.
 At this moment, according to Swift/UVOT data about evolution
of temperature and radius of the thermal component of the GRB/XRF\,060218 
afterglow, i.e. at a temperature of
kT $\sim$ (0.03---0.05)\,keV ($<$/$\sim $580000\,K) {\it at the moment}
($\sim 10^{4}$\,s) of the luminosity maximum,
 the size of the wind envelope
around the pre-SN must be equal to
   $>$/$\sim$300$R_{\odot}$
at the bolometric luminosity (4.6---35.5)x$10^{45}$\,erg/s.

4) {\it Velocity of the shock: }

    While the shock moves inside this stellar-wind envelope,
the radius corresponding to the thermal radiation component
(related to the shock breaking out from the center) is constantly increasing
from  $\approx$ 5.7$R_{\odot}$
to    $\approx$ 17$R_{\odot}$,
as it follows from the same data (Campana et al., 2006)
of the X-ray observations carried out with {\it Swift}/XRT
before maximum of the UV flash.
  The radius of the thermal component continues increasing
to $\approx$ 4700$R_{\odot}$
in 1.4 days (see Fig. 3 in Campana et al.), i.e. by the time of observation 
carried out with {\it Swift}/UVOT after the flash maximum.
  At this time the shock has already broke out to the ``surface''
of the stellar-wind envelope and luminosity of the thermal component
has decreased (see Fig.2).
  Dividing the path of the shock
(or the radius $\approx$3.29 +0.94/-0.93 x 10$^{14}$\,cm
of the thermal component at kT $\approx$ 3.7 +1.9/-0.9 eV)
by the time (1.4 days)
one can obtain expansion velocity of the photosphere related to the shock
by the moment of the spectral observations:
(2.7 +/- 0.8) x 10$^{9}$\,cm\,s$^{-1}$ (Blustin 2007).
  Such a velocity is typical of core-collapse SNe,
and it is comparable with line widths observed in the spectrum
of SN2006aj (Pian et al. 2006).

    Further it will be shown that the above estimates of the energy,
size and velocity obtained from the observations of the shock breakout effect
by the {\it Swift}/XRT/UVOT observatory are also confirmed when analyzing
our optical spectra of XRF\,060218/SN\,2006aj obtained in 2.55 and 3.55 days
after the beginning of the SN explosion,
i.e. when contribution of the {\it thermal} component of radiation
of the shock was still determining,
which is also indicated by strong blue excesses in our spectra (Fig. \,1).
  But, as is seen from the UBVR light curves (Fig.2),
in 5 days the GRB afterglow has been already noticeably reddening,
which is related to the change of the mechanism of the SN radiation
by this time, when the classical photospheric phase of the SN envelope
expansion begins.
This phase is usually observed in the core-collapse SNe and
it was described in reviews many times
(see, for example, Imshennik and Nadezhin, 1988).

    Section 2 describes early spectral observations of the
XRF\,060218/SN\,2006aj optical afterglow carried out with the 6-meter SAO 
RAS telescope.
  Section 3 describes the modelling of spectra with the
SYNOW code and discusses spectral manifestations of the wind envelope as 
wide absorptions near $5800\AA$ and $6100\AA$ (rest wavelengths with z=0) 
corresponding to the PCyg absorption in the $H\alpha$ hydrogen line.
  Section 4 contains the discussion of results in the context of: 1) an
obvious evolution of the $H\alpha$ traces by data for XRF\,060218/SN\,2006aj
obtained with BTA and other telescopes; 2) manifestations of the Colgate 
shock-breakout effect in the light curves, spectra, sizes and luminosities of
``usual'' Type Ib/c SNe; and 3) asymmetry of explosion of the core-collapse SNe.
  Conclusions are formulated in Section 5.
%

\section{Early spectral observations of the optical afterglow of 
  XRF\,060218/SN\,2006aj with the 6-meter SAO RAS telescope}

   Within the context of the program of follow-up observations of GRB optical
afterglows, the variable object detected with Swift/UVOT and associated 
with the XRF\,060218 event was spectroscopically observed with the 6-meter 
SAO RAS telescope.
   We used the Spectral Camera with Optical Reducer for Photometrical and
Interferometrical Observations (SCORPIO) mounted in the BTA primary focus
[$http://www.sao.ru/hq/lsfvo/devices/$ $scorpio/scorpio.html$].
   There were two sets of observations: at the epochs 16:31-17:18,
Feb. 20, 2006 UT and 16:31-17:16, Feb. 21, 2006 UT.
   The dispersing element was a combination of a transparent grid
and the prism VPHG550G with the spectral operational range and resolution
(FWHM - Full Width at the Half Maximum)
of $3500-7500$\AA\AA\ and $10$\AA, respectively.

   The processing of data was standard.
   It included subtraction of electronic zero, correction for flat field,
wavelength calibration with the help of comparison spectrum of a Ne-Ar lamp,
correction for atmosphere extinction and calibration by an absolute flux
with the use of observations of a spectrophotometric standard at every night.
   To detect a possible short-time variability and to control calibration
of absolute fluxes from night to night, a bright stellar-like object
was set at the input slit together with the optical transient
   (see Fig.~\ref{timur}).
   In all, 4 exposures were obtained in the first and second periods.
   We did not detect any short-time variability within these observational
periods, therefore we summed all 4 individual spectra to get a higher 
signal/noise ratio. 
    Figure~\ref{BTA_spectra}
shows average spectra in every night, 
    Table~\ref{meanbe}.
lists average epochs of these spectra.

    Then observational spectra (Fig.\,1) were corrected (Figs.\,5 and 6)
for galactic extinction according to the dust distribution maps of Schlegel,
Finkbeiner, \& Davis (1998).
   The value of $E(B-V)$ in the direction of the object
($\alpha_{2000}=03^h21^m39^s.683$,
 $\delta_{2000}=+16^{\circ}52^{\prime}01^{\prime \prime}.82$)
was equal to 0.14. 
   When accounting for extinction, the dust-screen model was accepted,
in which the expression for extinction has the form
$F_{int}(\lambda) = F_{obs}(\lambda)10^{0.4\cdot k(\lambda)E(B-V)}$,
where $F_{int}(\lambda)$ and $F_{obs}(\lambda)$ are issued
(without the extinction) and observed fluxes respectively.
   The extinction curve of the Milky Way ($k(\lambda)$) was taken
from the paper Cardelli, Clayton, \& Mathis (1989).


\section{The comparison of the observational spectra of XRF\,060218/SN\,2006aj
 with synthetic spectra}

    To interpret spectra we used the ``SYNOW'' code calculating synthetic
spectra of SNe (Branch et al. 2001).
   This multiparametric code was already applied many times
for a direct analysis of spectra of core-collapse SNe,
when hydrogen was detected in the phase of radioactive heating of the 
expanding SN envelope (Branch et al. 2002, Baron et al. 2005, Elmhamdi et 
al. 2006).
   The computation algorithm of model spectra is based on the
following assumptions: spherical symmetry, a homological expansion of layers
and {\it a sharp photosphere}.
   When interpreting early spectra, this optically-thick photosphere
emitting a continuum black-body spectrum
in some cases can be identified with the shock.
   At least this is valid in the first days (T$_{sp}$=1.95d and 2.55d from
Table 1) after XRF\,060218, when spectra (Fig.1) and photometry (Fig.2) 
still demonstrate a large blue excess, which is related with influence of 
the thermal component of radiation from the first short and powerful UV 
flash.
  It is assumed that spectral lines are formed over this expanding
photosphere as a result of resonance scattering which is interpreted in the 
SYNOW code in the Sobolev approximation (Sobolev 1958, Sobolev 1960, 
Gorbatsky and Minin 1963, Sobolev 1967).

    This code allows us to identify lines, to estimate velocity of the
photosphere expansion and the velocity range (or scatter) for lines of 
every ion detected in the spectrum.
  The detailed description of the SYNOW code can be found in the fore-quoted
papers Branch et al. 2001, Elmhamdi et al. 2006.
  To understand our results, the fact is essential that in the SYNOW
code the velocity of layers expansion {\it over the photosphere } is 
proportional to distance from every point of an expanding layer to the 
center ($v\sim r$).

    Judging by the form of observed spectral features, there are two cases:
the undetaching (``the undetached case'') and detaching (``the detached 
case'') of the layer, in which a spectral
line forms, from surface of the expanding sharp photosphere.
  If matter of moving gas layers located over the photosphere does not detach
from it (i.e. all layers radiate and screen photosphere starting from its 
level), then a line is observed in absorption or in emission like the case 
of here-considered case (Fig.~\ref{spektr})
of a line of the classic PCyg profile.
  We have chosen this model of layers undetached from photosphere
(``the undetached case'') to interpret our first spectrum 
(T$_{sp}$=2.55d, see Fig.\,1). 
   Besides, apparently, at this moment the expanding photosphere still can be
identified with the shock, because our first spectrum (with a larger excess
in its blue part) was obtained closer to the moment when the shock breaks 
out to the ``surface'' of the stellar-wind envelope surrounding the pre-SN 
--- the first maximum in Fig.~\ref{lightcurve3}.

    A layer detached from the photosphere manifests itself in spectrum as a
smoothed emission and a strongly blue-shifted absorption --- a ``remnant'' 
(part) of the PCyg profile --- as is shown in Fig.~\ref{spektr}.
  The ``detached case'' is realized when velocity of the layer
in which the line forms noticeably exceeds velocity of the photosphere.
  If $v\sim r$, then the absorption part of the PCyg profile is a result
of screening photosphere by a narrow layer; and the narrower is this layer
(radiating a spectral line), the closer is the emission to the photosphere
continuum.
  Such a model corresponds to our second spectrum
(T$_{sp}$=3.55d), at least, for HI lines which can be connected to the wind
envelope.
  It may be assumed that the shock provoked the motion and
``detachment'' of the uppermost layers of this extended wind envelope 
surrounding the GRB/SN progenitor star.

    In 3.55 days after the SN explosion, when the contribution of the thermal
component deceases (Fig.\,2) and the expanding envelope of SN becomes more 
and more transparent,
the shock cannot be already identified with the photosphere totally.
The so-called photosphere phase of the SN explosion begins.
  The SYNOW code was applied mostly to interpret spectra of this
phase (Elmhamdi et al. 2006, see also references ibidem).

     In Figs.\,\ref{undetached} and \ref{figure1c}
the BTA spectra of XRF\,060218/SN2006aj, obtained in 2.55 and 3.55 days
after the event are compared to synthetic spectra modeled with the SYNOW code.
 The first spectrum taken right after the bright UV burst
 (Fig.\,\ref{lightcurve3}),
is easily modeled for a rather simple set of parameters. 
  We have chosen temperature of the photosphere ($t_{bb}$) used in fitting
the energy distribution in the spectrum in Fig.\,\ref{undetached} (rest 
wavelengths) to be equal to 9000\,K.
  This corresponds also to the temperature fall rate by results
of Swift/XRT/UVOT observations (Campana et al. 2006), according to which
by the time of our spectral observation the temperature must be
less than 10000\,K.
  In the spectrum taken in 2.55 days the SYNOW model shows the photosphere
velocity equal to 33000\,km\,s$^{-1}$.
  This parameter is within errors of estimates of the photosphere expansion
velocity connected with the shock, which is obtained from the same 
Swift/XRT/UVOT data ((2.7 +/- 0.8)x10$^{9}$\,cm\,s$^{-1}$).

\begin{figure*}[p]
\setcaptionmargin{5mm}
\onelinecaptionsfalse
  \includegraphics[width=15cm]{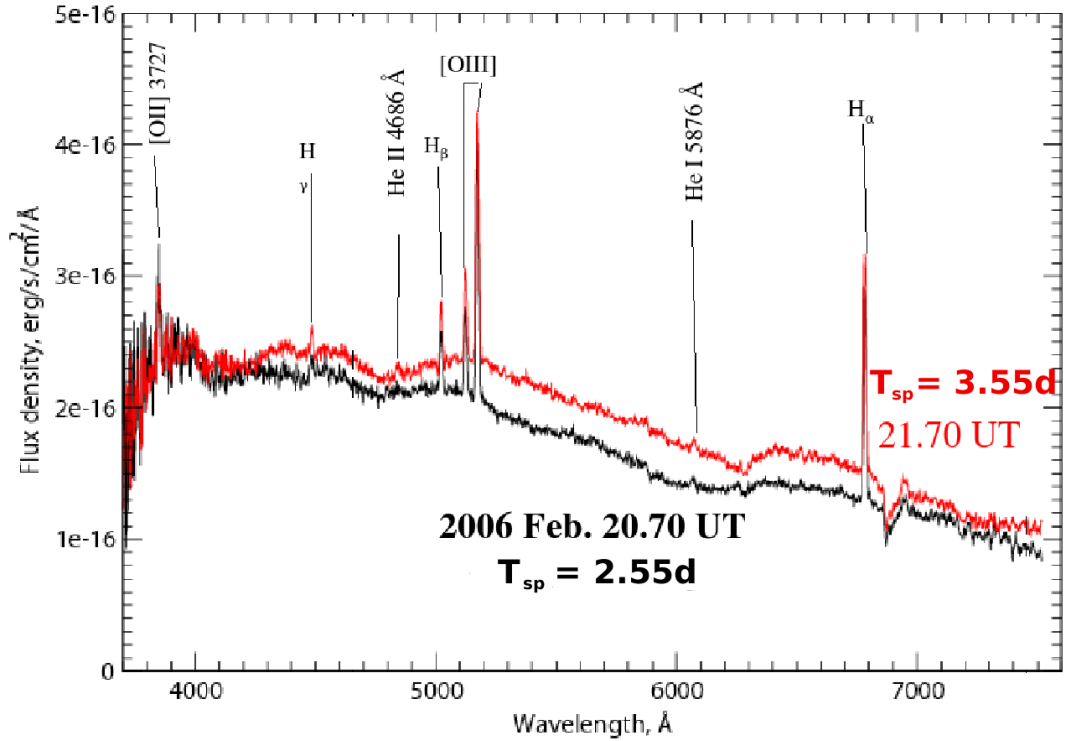}
\captionstyle{normal}
\caption{Observational spectra of the XRF\,060218/SN\,2006aj afterglow
obtained with the BTA (Fatkhullin et al. 2006).
UT times and observational times after {\it the beginning} of the
SN explosion are indicated.
Emission lines of the host galaxy (z=0.0331) are also shown.
}
\label{BTA_spectra}
\end{figure*}

    This means that by the moment of our spectral observations and
approximately in a day after the last Swift/UVOT observation of the fading 
UV flash (the shock breakout), the shock velocity remained within the same 
limits.
 Furthermore, a wide and hardly noticeable depression in continuum
(see Fig.~\ref{BTA_spectra} with observed wavelengths) at 
$\approx 5900 - 6300\AA$, and an almost unnoticeable excess of flux within 
the wavelength range $\approx 6300 - 6900\AA$ is best fitted by a wide 
$H\alpha$ PCyg profile (see Fig. ~\ref{undetached}) at the same velocity 
33000\,km\,s$^{-1}$.
   In the SYNOW code this corresponds to ``the undetached
case'' in Fig.\,\ref{spektr}.

    It may be supposed that at this time {\it a part} of the stellar-wind
envelope located over the photosphere or an extended layer over it moves 
together with the photosphere (the shock), and we observe acceleration or 
detachment of upper layers of this envelope.
   This is one more result
(in $\sim $2 days after the bright UV (UVOT) flash) of the shock breakout to
``the surface'' of the massive circumstellar envelope which was (almost)
resting before the SN explosion.
  According to estimates given in Introduction,
the size of this envelope was
	 $>$/$\sim$300\,$R_{\odot}$,
at least at the moment of the shock breakout in $\simeq$11 hours
after the event.
  To demonstrate that an almost unnoticeable oscillation of flux in
Fig.\,\ref{BTA_spectra} within the range $\approx 5900 - 6900\AA$ of the 
observed spectrum may be indeed a very wide $H\alpha$ PCyg profile
for velocities about 0.3x10$^{10}$\,cm\,s$^{-1}$,
in Fig.\,\ref{undetached} we give a
model spectrum with the classic PCyg profile, but for a much less velocity 
equal to 8000 km s$^{-1}$.
  (The laboratory $H\alpha$ wavelength is marked by a narrow emission
of the host galaxy.)

     We made fitting of this undetached case by different synthetic spectra,
but velocity of the photosphere ($V_{phot}$), all elements and their ions
(vmine) {\it was identical} and equal to 33000 km s$^{-1}$.
  Basic parameters of calculation are given in Table 2 for the
``undetached case'' in Fig.~\ref{undetached}.
   The only difference is in effective depths of ion line formation
$\tau$ (tau1), which can be chosen relatively small ($\tau < 0.5$) for all
elements and their ions, for the observed spectrum to be described
satisfactorily.
   Within the wavelength range $\approx 4500 - 7200\AA$ it is hydrogen that
has the largest value $\tau$=0.2.
   Other elements and their ions were taken into account in the synthetic
spectrum with even smaller $\tau$, which may reflect small variations
in relative abundance of other elements in comparison with the most abundant
one.
  This may be connected with the fact that in this first spectrum
($T_{sp}$=2.55 days) the contribution of the stellar-wind envelope is still
determinant.

  It should be said that we were trying to get the best agreement between
the observed and synthetic spectra in their range where observations give the
best signal/noise ratio and the theoretic spectrum is within the grass
or in its nearest proximity.
    In particular, the outlining deficit of flux to the left of the emission
line of the host galaxy [OII]\,$3727\AA$ (in Fig.\,\ref{undetached})
can be described by influence of CaII with $\tau$=0.5.
      This is the largest values of the effective depth of formation
that we used in this paper dedicated to interpretation of the earliest 
spectra of the XRF\,060218/SN\,2006aj afterglow.

  In our second spectrum for $T_{sp}$=3.55 days (see Fig.\,\ref{BTA_spectra}
with observed wavelengths) we see some absorption at $6300\AA$ interpreted 
here as a strongly blue-shifted ``remnant'' of the $H\alpha$ PCyg profile. 
   The best fitting of the synthetic spectrum to the observed one corresponds
to the case denoted as ``the detached case'' in Fig.\,\ref{spektr}, when a 
part of the wind envelope has already detached from the photosphere.
      I.e. contribution of the thermal burst (Fig.2) is quickly decreasing
and the shock is becoming more and more transparent --- the photosphere
phase of the SN envelope expansion begins (Branch et al. 2001).

   We were modelling this second spectrum within the velocity range
18000 km s$^{-1} < v < $ 24000\,km\,s$^{-1}$ (see Fig.\,\ref{figure1c} of
rest wavelengths with z=0).
    The synthetic spectrum which is the closest to the observed one is the
spectrum with parameters of the model from Table 3,
where the photosphere velocity is equal to 18000\,km\,s$^{-1}$, and its 
temperature is t$_{bb}$ = 8200\,K.
  In the synthetic spectrum we took into consideration the contribution
of lines with small $\tau (< 0.3)$ of such elements as
HI, HeI, FeII, SiII, OI, CaI, CaII, TiII, NI, CI, CII,
MgI, MgII, NaI.
   Fig.\,\ref{figure1c} shows location of some of these lines in those
spectral regions where the contribution of the ion into the spectrum is 
essential with given parameters of the model from Table 3.
   The list of lines used in the SYNOW code (``the reference lines'')
is given in Elmhamdi et al. (2006), detailed information is given directly
in the SYNOW code
{http://www.nhn.ou.edu/$\sim$parrent/synow.html}.

     A wide absorption with minimum near $6100\AA$ (Fig.\,\ref{figure1c})
can be described by influence of HI for ``the detached case''
(Fig.\,\ref{spektr}) at $\tau = 0.16$.
    In this region ($\approx 5700 - 6500\AA$ in Fig.\,\ref{figure1c}) the
synthetic spectrum is characterized by a smoothed emission from the red 
side and a strongly blue-shifted absorption with minimum near $6100\AA$ ---
``the remnant'' of the $H\alpha$ PCyg profile of the line HI.

    Thus, in $T_{sp}$=3.55 days, hydrogen has already detached from the
photosphere and the corresponding layer moves at a velocity of 
24000\,km\,s$^{-1}$ (see Table 3).
   Here also all ions were taken with the small values $\tau < 0.3$.
But we failed to describe totally this observed spectrum
with identical values of all other parameters,
as was made in Table 2 (``the undetached case'').
   The layers where lines of other elements mainly
form either move at the velocity identical to that of the photosphere 
(18000\,km\,s$^{-1}$ for the lines of OI, CaI, CaII, TiII, NI, NaI),
or at the velocity of hydrogen (24000\,km s$^{-1}$ for the lines of HI,
HeI, FeII, SiII, CI, CII, MgI, MgII).
   For the best description of the observed spectrum,
typical velocity values (the parameter ve in Table 3)
determining the characteristic thickness of layers occupied by every 
element should also be chosen different.
   It may be assumed that this
spectrum is already affected more by the chemical composition of the
scattering SN envelope (or envelopes) which changed as a result of evolution
and explosion of a far-evolving massive core of the progenitor star.

\section{Discussion of the results }

{\bf 4.1 Evolution of spectrum of SN\,2006aj
and other core-collapse SNe.}

\begin{figure*}[p]
\setcaptionmargin{5mm}
\onelinecaptionsfalse
   \includegraphics[width=15cm]{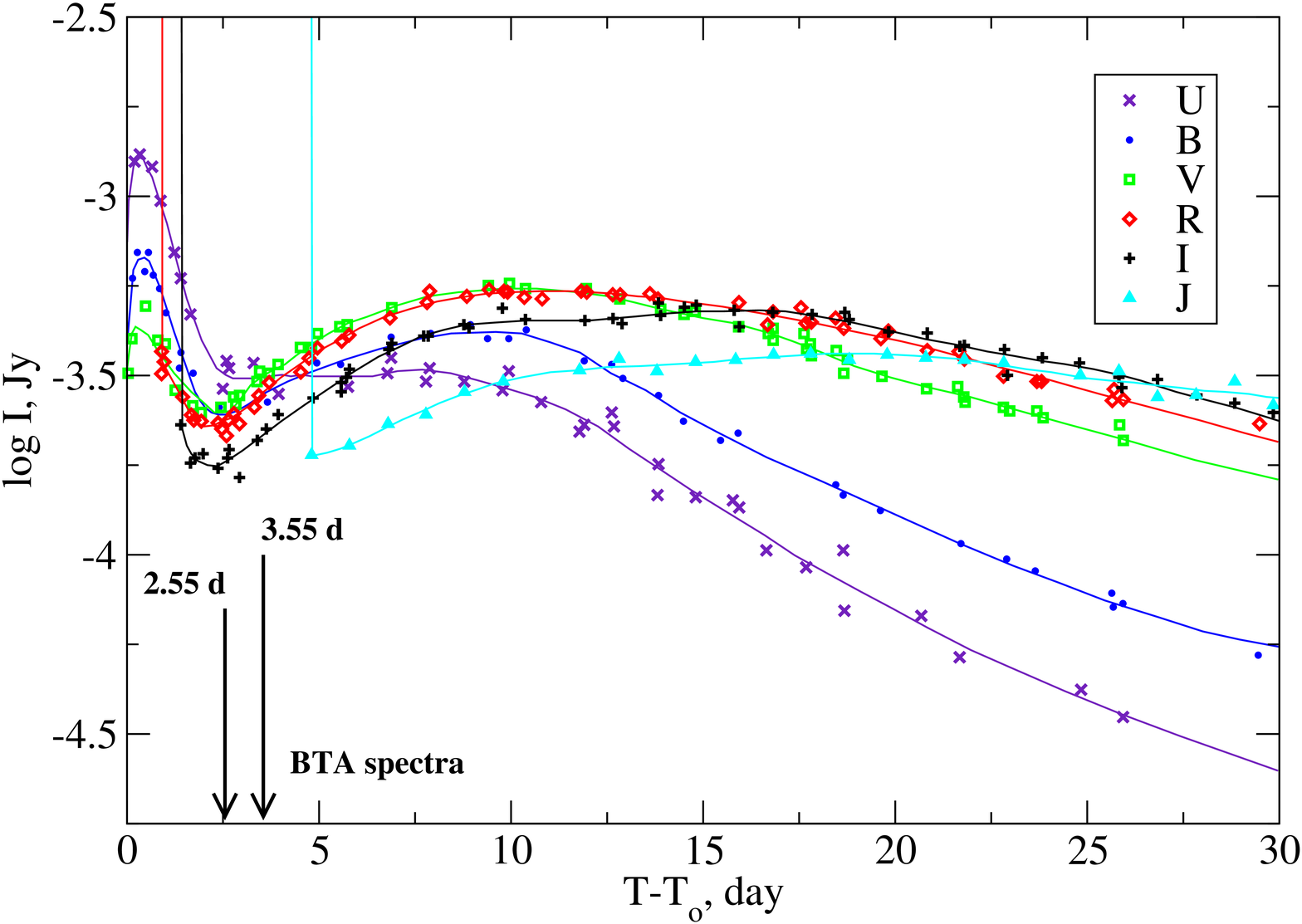}
\captionstyle{normal}
\caption{
Optical (U, B, V, R, I) and infrared (J)
light curves of XRF\,060218/SN\,2006aj (Jelinek et al. 2007).
The first maximum corresponds to the UV flash/burst --- the shock breakout
effect (see the text).
Spectra from Table 1 refer to the transitional region near minimum of the 
curves. 
Arrows point times of BTA spectra after the beginning of SN explosion. 
T$_o$(days) corresponds to 2006 Feb. 18.149 UT.
}
\label{lightcurve3}
\end{figure*}

\begin{figure*}[p]
\setcaptionmargin{5mm}
\onelinecaptionsfalse
   \includegraphics[width=9cm]{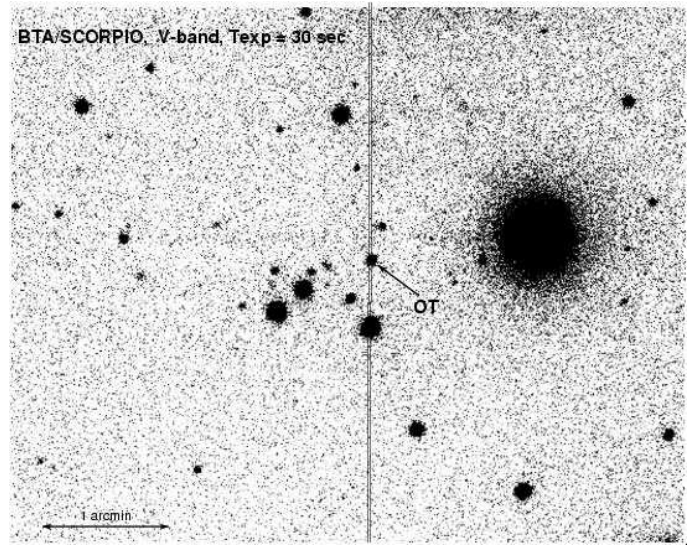}
\captionstyle{normal}
\caption{Field of the optical tranisent. BTA optical observations started 
on Feb. 20.647, 2006 UT
 ($\approx $ 60 hours after the GRB event)
 P.A. = $3^\circ$ is a position angle of slit,
$V_{OT} = 18.16$, 
$(B-R)_{OT} = 0.3$}
\label{timur}
\end{figure*}

\begin{figure*}[p]
\setcaptionmargin{5mm}
\onelinecaptionsfalse
  \includegraphics[width=15cm]{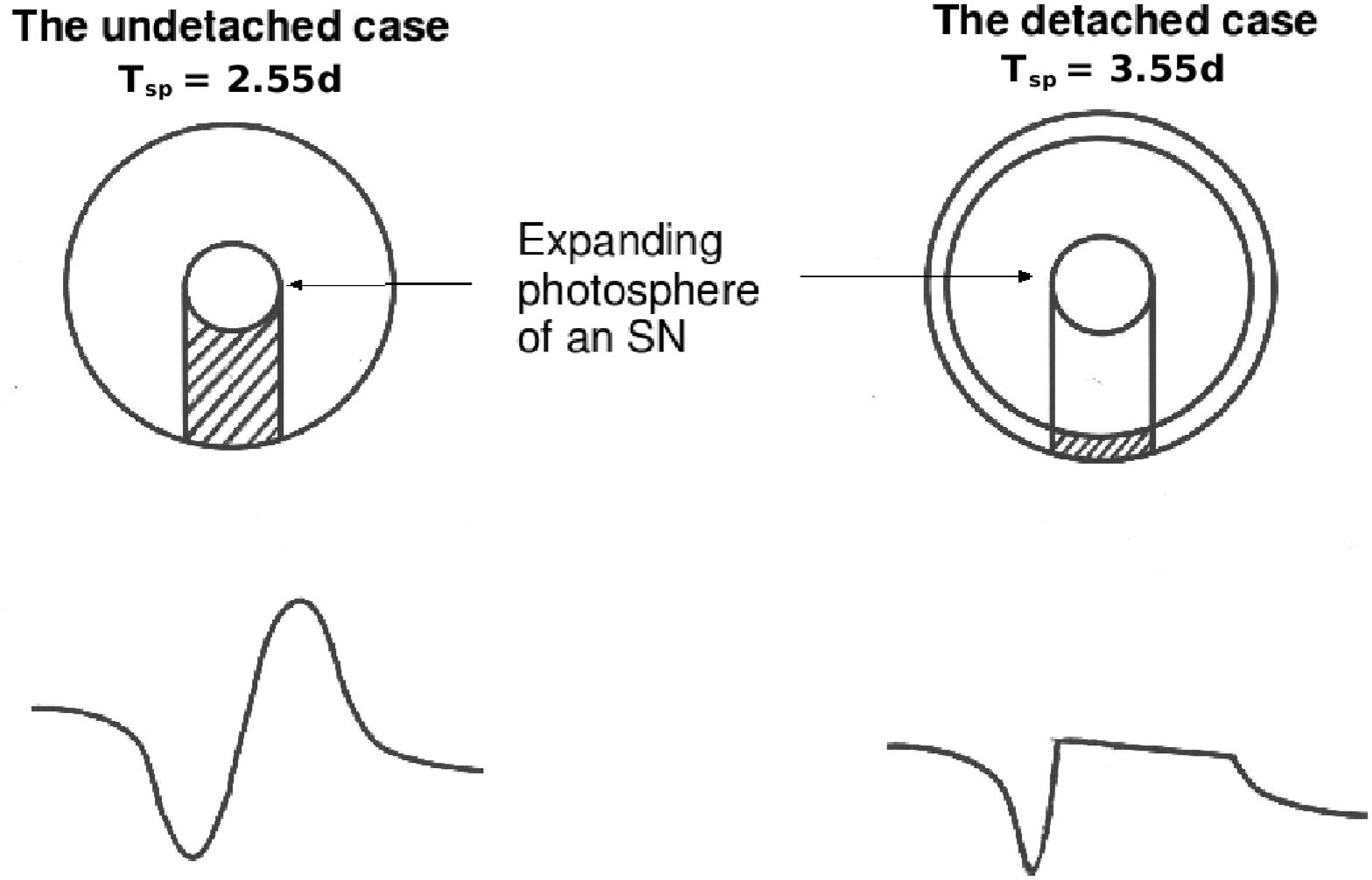}
\captionstyle{normal}
\caption{Line profiles corresponding to the cases when envelope layers (i.e. 
layers {\it over} the photosphere) detach or do not detach from the 
expanding photosphere, when the gas expansion velocity increases 
proportionally to distance to the center ($v\sim r$, see the text). The 
shaded regions form the absorption component of the PCyg profile.}
\label{spektr}
\end{figure*}

\begin{figure*}[p]
\setcaptionmargin{5mm}
\onelinecaptionsfalse
  \includegraphics[width=15cm]{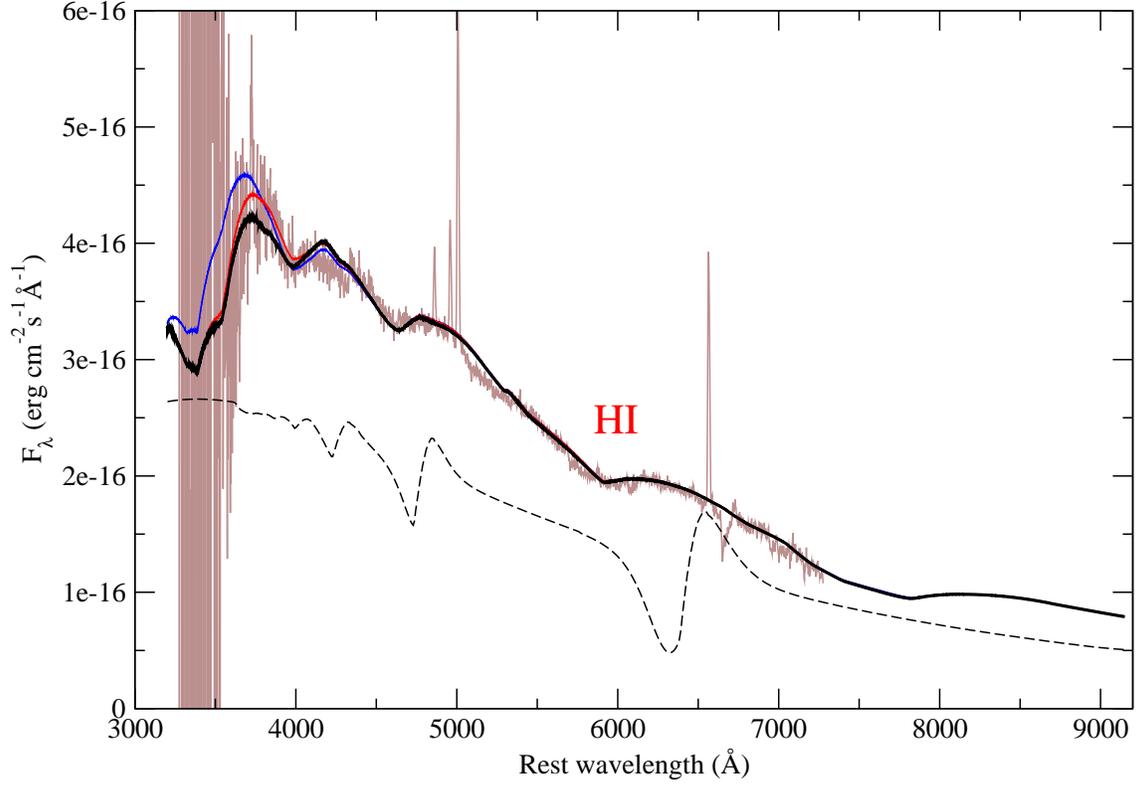}
\captionstyle{normal}
\caption{
  The spectrum of the XRF/GRB060218/SN2006aj afterglow in rest wavelengths (z=0)
obtained with BTA in 2.55 days and corrected for galactic extinction.
  Its fitting for ``the undetached case'' by synthetic spectra with
the velocity of the photosphere ($V_{phot}$), all elements and their ions
equal to 33000 km s$^{-1}$ is shown by smooth lines differing only
in the blue range of the spectrum at $\lambda < 4000\AA$ (see the text).
  Main parameters of calculation of the synthetic spectrum represented
by the thick black line are given in Table II.
  HI denotes the $H\alpha$ PCyg profile at $V_{phot}$ = 33000 km s$^{-1}$.
  The model spectrum for the photosphere velocity 8000 km s$^{-1}$ is shown
for example by the dashed line as an example of the $H\alpha$ PCyg
profile.
}
\label{undetached}
\end{figure*}

\begin{figure*}[p]
\setcaptionmargin{5mm}
\onelinecaptionsfalse
  \includegraphics[width=15cm]{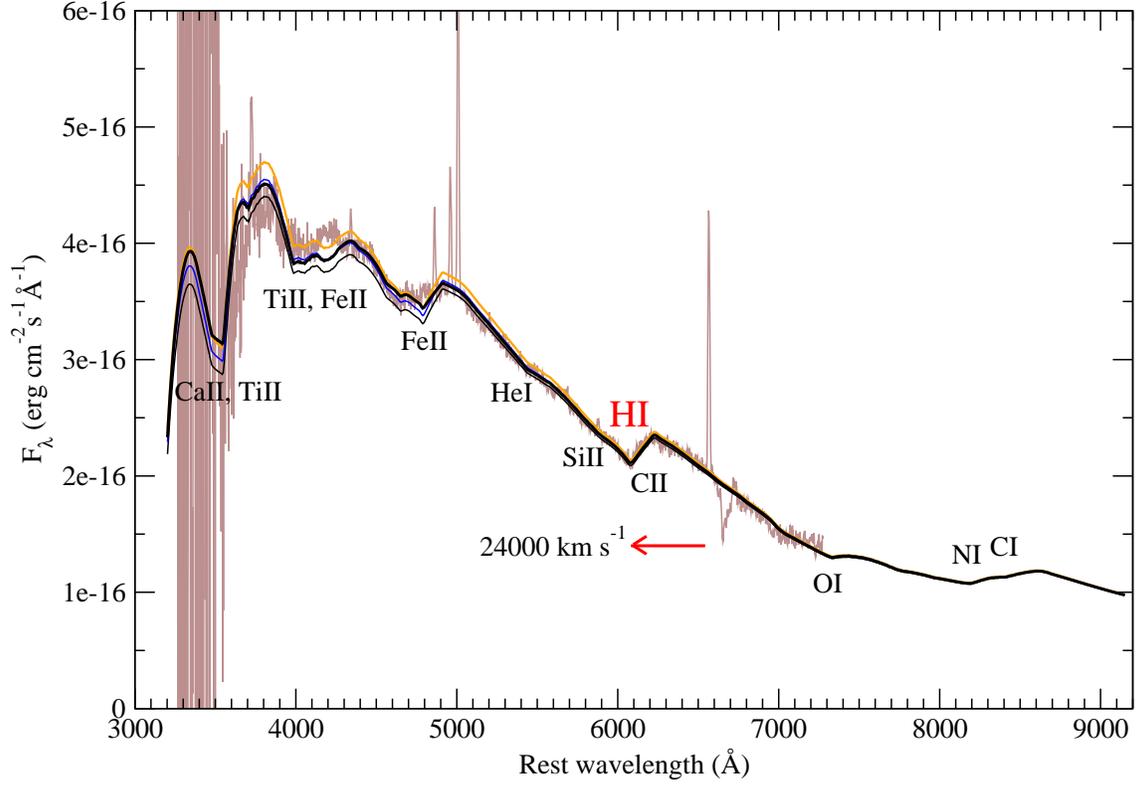}
\captionstyle{normal}
\caption{
  The spectrum of SN2006aj/XRF060218 (rest wavelength at z=0) obtained with
BTA in 3.55 days and corrected for galactic extinction.
  Synthetic spectra are shown by smooth lines.
  Locations of spectral lines of some ions and blends of their lines
are shown in those parts of the spectrum where contribution of this ion
into the spectrum is essential for given model parameters.
  The thick black line is the synthetic spectrum with parameters from
Table III at which the absorption with minimum about $6100\AA$ is described
by suppressing influence of HI for ``the detached case''.
  This is a strongly blue-shifted part of the $H\alpha$ PCyg profile
at the velocity of expansion of the detached HI layer equal to 
 24000\,$km\,s^{-1}$.
}
\label{figure1c}
\end{figure*}

  A wide and low-contrast feature in the spectrum of the XRF060218/SN2006aj
afterglow with minimum at $\approx $$6100\AA$ (in rest wavelengths) can
be traced in all early spectra obtained with our and other telescopes 
starting from Feb 21.70 UT (see Table 1).
  The spectrum which is the closest in time (Feb 21.93 UT) to our second
spectrum from Table 1 is the spectrum obtained with NOT
(see Fig.\,2 in Sollerman et al. 2006), which also contains minimum
at the same wavelength $\approx6100\AA$.
  The same feature is also seen in the ESO Lick spectrum obtained on
Feb 22.159 UT, but its signal/noise ratio (it is the second spectrum in 
Fig.\,1 in Mazzali et. al. 2006) is noticeably less than for 
two above-mentioned spectra.
  From data adduced in Fig.1 in the paper Mazzali et. al. (2006) it is seen
that this feature obviously evolves starting from the very first VLT spectrum
and becomes the deepest in the spectrum obtained with VLT on Feb 23.026 UT,
i.e. in about 5 days after XRF\,060218.
  Later spectra obtained after minimum in the light curve
(see our Fig.\,\ref{lightcurve3}) are more and more affected by the
increasing wide absorption {\it related to } SiII {\it near} $6000\AA$,
as was noted by Mazzali et. al. (2006).
  But the narrow minimum at $6100\AA$ is still seen even in the March's VLT
spectra.
  This is very similar to what is observed in spectra of some Type Ib SNe
(Parrent et al. 2007), where the narrow absorption related to $H\alpha$
is also detected and evolves in the spectra obtained in the beginning of
the long rise identical to that in our Fig.\,\ref{lightcurve3} to the
typical (of Type Ib and Ic SNe) maximum of brightness.

  The start of evolution of the spectral feature at $6100\AA$ which is then
traced for XRF\,060218/SN\,2006aj in the data from other telescopes is also
confirmed in our above-mentioned spectra.
  But we also include here the first BTA spectrum of Feb 20.70 interpreting
the wide and almost unnoticeable depression of continuum within the range
$5600 - 6600\AA$ ($5800 - 6800\AA$ are observed wavelengths in
Fig.\,\ref{BTA_spectra} for $T_{sp}$=2.55d) as the $H\alpha$ PCyg profile
for velocities of 33000\,km\,s$^{-1}$.
  The same small oscillation of continuum can be seen also in the very first
VLT spectrum of Feb. 21.041 UT (see the same Fig. 1 in Mazzali et. al. 2006)
obtained in 8 hours after our first spectrum.
  As to the MDM spectrum obtained in 14.5 hours before us (see Table 1),
here also one can see a weak change of continuum at $\sim5700\AA$ which is
noticeable even at the low signal/noise ratio (Mirabel et al. 2006).
  Here also the same very wide $H\alpha$ PCyg profile for velocities of
$\sim$30000\,km\,s$^{-1}$ is quite competitive to the identification
suggested by these authors.
  Thus, taking into account all early observations from Table 1, it may be
said that we do observe evolution of the optical spectrum --- {\it a transition}
from the phase of the Colgate shock breakout effect and the powerful
(thermal) burst related to it to spectra of the phase of the SN brightness
increasing which corresponds to radioactive (non-thermal) heating at the
decay $^{56}$Ni$\to^{56}$Co$\to^{56}$Fe.

  Maybe, this is the main difference and novelty of our approach to
interpretation of early spectra of XRF\,060218/SN\,2006aj to that 
represented in the paper by Mazzali et. al. (2006), where the VLT/Lick 
spectra of this SN were identified and where the question is {\it only} on 
theoretic spectra for the phase of radioactive heating --- see the light 
curve (Fig.\,2) in their paper.
  And when analyzing spectra of usual Type Ib-c SNe
(``stripped-envelope SNe''), the authors of the SYNOW code
(Branch et al. 2001, Branch et al. 2002, Baron et al. 2005, Elmhamdi et al. 
2006) were calculating their synthetic spectra with traces of the $H\alpha$ 
line, but yet not for so early phases with a considerable contribution of 
the thermal radiation from the UV flare and so large expansion velocities 
which were observed for XRF\,060218/SN\,2006aj.

  The evolution of all 16 ESO Lick and ESO VLT spectra of SN2006aj
(Pian et. al. 2006) was modeled up to March 10, 2006
(20 days after XRF\,020618) in the fore-quoted paper by Mazzali et al. (2006)
with the help of a more sophisticated code of synthesis of SN spectra by the
Monte Carlo method based under the same assumptions as the SYNOW code,
but with consideration of model distributions of density and temperature
in the envelope over the photosphere, radiation transfer in terms of
transitions in lines and electronic scatting (details of the method
are expounded in the papers Mazzali and Lucy (1993), Lucy (1999),
Branch et al. (2003).
 All typical signs of Type Ic SNe increases in spectra before the wide
maximum in the SN\,2006aj light curves ($\approx$10 days after GRB in 
Fig.\,\ref{lightcurve3}) and after it.
 The spectra were modeled for the velocity range
20000\,km\,s$^{-1} < v <$ 30000\,km\,s$^{-1}$.
 The strongest features mentioned by the authors are the lines of
FeII, TiII,
and at later phases --- CaII ($<$ 4500$\AA$), FeIII и FeII (near
5000$\AA$), SiII (near $6000\AA$), OI (near 7500$\AA$) and CaII (near
8000$\AA$).

  We also took into consideration all this list of lines when interpreting
our early spectra with the SYNOW code (see Fig.\,\ref{figure1c}) for almost 
identical range of velocities, but we also included contributions of the 
H\,I и He\,I lines.
  From calculations adduced in Mazzali et al. (2006) it
is seen that in the very early spectra of Feb 21, Feb 22, Feb 23 the 
authors did not take into consideration abvious traces of the $H\alpha$ 
line, and in later spectra only the increasing influence of 
Si\,II near $6000\AA$ was considered --- see Fig.\,1 in Mazzali et al. 
(2006) and the caption to it.
  At the same time, as was mentioned above, the
absorption with minimum at $\simeq6100\AA$ can be traced in the 
{\it observed} spectra up to March,4.
  It is the most noticeable feature
near $6000\AA$ in the early spectrum of Feb.23 in the whole range from 
$\sim5000\AA$ to $\sim8500\AA$, where the Monte Carlo method does not fix 
any absorption at $\simeq6100\AA$.

{\bf 4.2 The signs of hydrogen in core-collapse SNe spectra.}

  The signs of hydrogen in spectra of Type Ib and Ic (Ib-c) SNe are no news.
 In particular, the signs of hydrogen and evolution of the blueshofted
$H\alpha$ line were already found with the help of the same SYNOW code in 
the analysis of a time series of optical spectra for usual core-collapse 
Type Ic and Ib SNe.
  In this analysis, special attention addressed to traces of hydrogen in
observations of especially these stripped-envelope Type Ib-c SNe.
  Though, by (formal) definition, the Type Ic and Ib supernovae does not have
conspicuous lines of hydrogen  in its optical spectra.
  The Ib-c SNe are usually modelled in terms of the gravitational
collapse of massive and bare carbon-oxygen cores which stripped envelope
before collapse, and, apparently, signs of this envelope must be present
{\it always} in spectra of these SNe as hydrogen lines.
  Though most often hydrogen can be identified more or less reliably only
in sufficiently early spectra of Ib-c SNe, as was shown in
Branch et al. 2002, Baron et. al. 2005,
Branch et al. 2006, Elmhamdi et al. 2006, Parrent et al. 2007.

   A competitive hypothesis for description of the spectral feature with
minimum near $6100\AA$ may be its interpretation as an absorption component
of the PCyg profile of the line CII $6580\AA$ (Branch et al. 2006).
   But if the whole extended  ($>$/$\sim$300\,$R_{\odot}$)
stellar-wind envelope was seen at first in X-rays and subsequently
as a UV flare in the XRF\,060218/SN\,2006aj afterglow, then it is hydrogen
that should be connected at least to that part of the relic envelope
which did not evolve
and refers to the stellar-wind stage of evolution of the core-collapse
progenitor star, and which originated long before the SN explosion.
   The fact that it is hydrogen that always shows
the largest contrast of velocity between the HI layer and the
photosphere in comparison with other elements in spectra of Type Ib-c
core-sollapse or ``stripped-envelope'' SNe (Branch et al. 2002 (see Fig. 23),
Branch et al. 2001 (см. Fig. 9), Elmhamdi et al. 2006 (see Fig. 5)) also is 
in favour of the idea that it is these layers related to the wind envelope 
that are the first to come into motion as a result of SN explosion.
   One can estimate also the mass of
this moving part of the envelope (the mass of gas in the $H\alpha$ or HI 
layer) with the help of the equation from the paper Elmhamdi et al. (2006) 
in the following way:

      M($M_{\odot}$) $\simeq$ (2.38 x 10$^{-
5}$)v$_{4}$$^{3}$t$_{d}$$^{2}$$\tau$($H\alpha$) \ \,,

where time after the explosion t$_{d}$ is in days, the velocity of layer 
v$_{4}$ is in units of 10$^{4}$\,km\,s$^{-1}$, and $\tau$($H\alpha$) is the 
depth of line formation.
   In this equation deduced for the Sobolev optical
depth in the expanding envelope (Castor 1970; Elmhamdi et al. 2006), we 
used the velocity 24000\,km\,s$^{-1}$ which refers only to that part of the 
HI layer which came into motion (see data in Table 3).
   This corresponds to the \textit{detached} case in Fig.\,\ref{spektr}.
   As a result, the mass of the HI layer turns out to be
$\sim$0.0006$M_{\odot}$, and its distance to the center by this
time --- 3.55 days after the beginning of the SN explosion --- is in
any case not less than 7.36 x 10$^{14}$cm.

{\bf 4.3 The Colgate shock-breakout effect in XRF\,060218/SN\,2006aj
and in other SNe --- light curves, spectra, luminosities and sizes.}

  The Type Ib and Ic SNe have long been observed, it is has already long been
considered that their progenitors are most probably the WR stars surrounded 
by more or less dense wind envelope which resulted from evolution of a 
core-collapse star.
  The shock arising in explosion of evolved star core passes through
the envelope and leads to a bright and short X-ray and UV flash
which can last several hours --- duration of the flare depends on how
massive and extended was the wind envelope surrounding the progenitor star 
before the SN explosion.
  Interaction between the shock caused by the SN explosion and the envelope
(the shock breakout effect) has also been predicted and comprehended long ago
(Colgate 1968, Bisnovatyi-Kogan et al. 1975, Blinnikov et al., 2002).
  In particular, the paper by Calzavara and Matzner (2004) is dedicated to
future systematic observations of this effect.
  But so far, before the event XRF\,060218/SN\,2006aj, observations
of the almost total effect has been gained for only small amount of
the core-collapse SNe.

    The shock breakout effect which was very short due to the compactness of
the blue supergiant   (20-30$R_{\odot}$)
was observed (though not from the very beginning) in the famous Type II
SN\,1987A: there are quite a few data and the physics is described rather
well --- see the large review by Imshennik and Nadezhin (1988).
   During a long time (2-3 weeks), one could
observe the shock breakout effect in a very extended ($\sim 10^{15}$\,cm) 
envelope of Type IIn SN\,1994W (see, for example, in Chugai et al. (2004)).
   In other cases of the Type Ib-c SNe the shock breakout effect was observed
only in its very ending, before the subsequent increasing of the SN 
brightness corresponding to the radioactive heating  
$^{56}$Ni$\to^{56}$Co$\to^{56}$Fe --- for SN\,1999ex, Stritzinger et al.
(2002).
   Identical ``remnants'' of the shock breakout effect were seen (in
the R band) even in SN\,1998bw, which is usually related to GRB\,980425
(see Galama et al., 1998).

\begin{figure*}[p]
\setcaptionmargin{5mm}
\onelinecaptionsfalse
  \includegraphics[width=15cm]{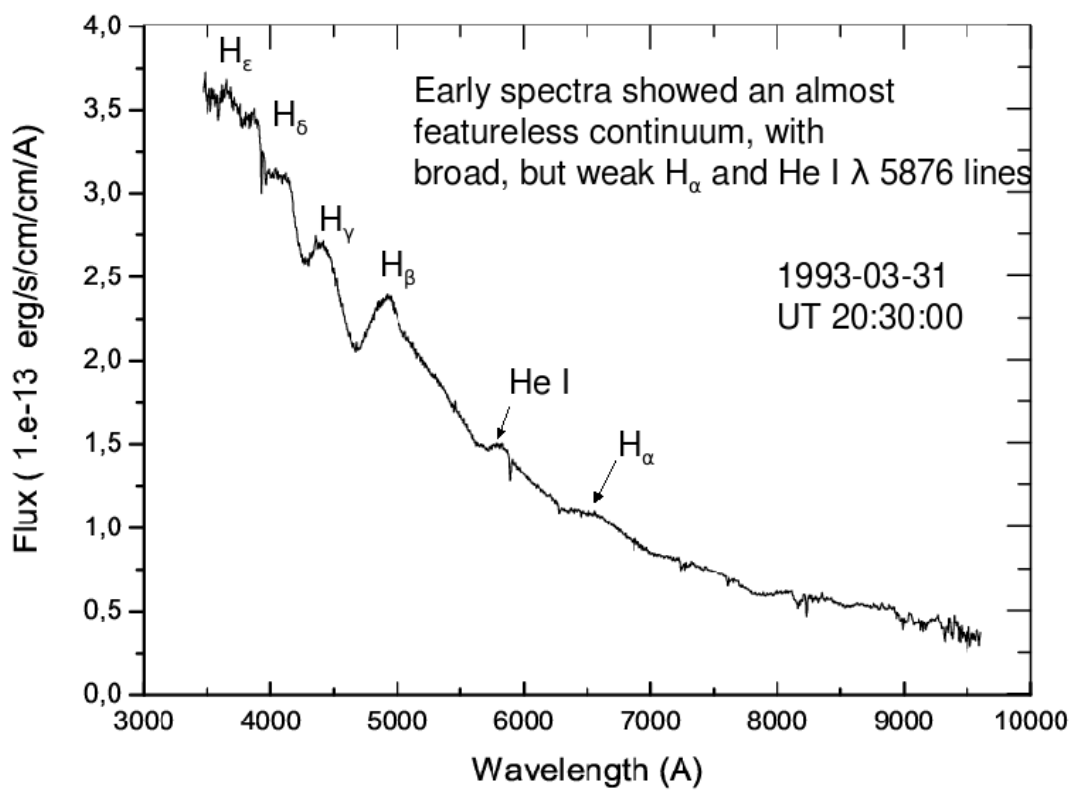}
\captionstyle{normal}
\caption{ The earliest optical spectrum of SN\,1993J (see the text).
The spectrum is taken from the database SUSPECT - The Online Supernova 
Spectrum Database 
http://bruford.nhn.ou.edu/suspect/ (Richardson et. al. 2002)
}
\label{sn1993jspectra}
\end{figure*}

{\it SN\,1993J in M81:}
    It can be said that SN\,1993J was observed almost in the very beginning of
explosion.
   ``Almost'' because in this case the moment of the SN explosion
beginning is known with an accuracy of $\simeq$12 hours, but not to seconds 
as in XRF/GRB SNe.
    Initially, this SN was classified by spectrum as the
Type II SN because of appearance of hydrogen lines in its early spectra
(Fig.\,\ref{sn1993jspectra}), but after a time, the SN type changed to Ib 
(Flippenko, Matheson and Ho, 1993), when hydrogen in spectra stopped being 
detected reliably.
   Early spectra of SN\,1993J in Fig.\,\ref{sn1993jspectra}
show a strong UV excess typical of the shock breakout effect and a smoothed 
comtinuum almost without lines above $5000\AA$, similar to the spectrum of 
XRF\,060218/SN\,2006aj in Fig.\,\ref{undetached}, though in the case of 
SN\,2006aj the expansion velocities are much higher (see below the
remarks about possible asymmetry of explosion of SNe identified with GRBs). 
   Besides, SN\,1993J had an unusual light curve, when luminosity quickly
increased to the first maximum and then headily fell during several days 
and then is has been slowly increasing for the second time during the 
subsequent two weeks.
   The behavior is analogous to that shown in Fig.\,\ref{lightcurve3}
for XRF\,060218/SN\,2006aj.
   The SN\,1993J light curve has been long and well modeled by many groups:
Nomoto et al. (1993), Young et al. (1995), Shigeyama et al. (1994).
   These papers say that the shock breakout in SN\,1993J is naturally
explained by interaction between the shock and the extended
($300 R_\odot \simeq 2.1\cdot 10^{13}$\,cm)
hydrogen envelope of mass $\sim 1 M_\odot$ around the progenitor star.
   Here the luminosity achieved in the peak of the first/quick maximum
which lasts only 4-5 hours can achieve a value of $\sim 10^{45}$\,erg/s
with the total photon energy release $\simeq5\cdot 10^{49}$\,erg.

   No wonder that approximately identical parameters of the envelope explain
the shock breakout burst in the case of XRF/GRB\,060218/SN\,2006aj.
   Here also the total energy turns out to be of the same order of
$\sim 10^{49}$\,erg.
   But in the case of SN\,1993J there was no gamma-ray burst, and,
most probably, this is caused by asymmetry of the SN explosion
(see below about that).

    If GRB/XRF\,060218/SN\,2006aj is another case when the Colgate shock
breakout effect was observed in pure form {\it from the very beginning} of 
the SN explosion , then it may be understood as quite a definite hint that 
the GRB itself is the first signal in gamma-rays that the
collapse of a massive core started, which is followed by the total
process --- an anti-collapse and explosion of a massive SN: after the GRB
an X-ray flash (XRF) and then a powerful UV flash can be observed.
   Then the key moment in optical opbservations of transient sources related
to XRF/GRBs can be {\it the search} for all manifestations of wind 
envelopes around core-collapse progenitor stars in early spectra and in 
photometry of the XRF/GRB afterglows, because interaction of the SN shock 
and these envelopes is the very first event in optical after the beginning 
of collapse of the massive stellar core.
   Here one may expect observing new effects related, for example,
to the same asymmetry of explosion.

{\bf 4.4 Asymmetry of the Type Ib and Ic SNe explosions. }

    The burst XRF\,060218/SN\,2006aj was a classical XRF event indeed
(Heise et al. 2001, Campana et al., 2006).
  The fact that in the case of usual and nearby SNe the explosion does not
begin with a GRB is naturally explained by an aymmetric,
axial-symmetric or bipolar (with formation of jets) explosion of the
core-collapse SNe.
  Now one of the most popular conceptions (see references in the paper by
Soderberg et al., 2005) proceeds from the idea that in the case of
flashes of the XRF type an observer is out of the beam in which the most 
gamma-ray radiation is concentrated for one reason or another.

   The farther is an observer from the SN explosion axis, the more of X-ray
radiation and the less gamma-ray quanta are in the spectrum of the 
flash --- GRBs transform to X-ray Rich GRBs (like GRB\,030329) and become 
X-ray Flashes (Sokolov et al. 2006).
   When observing at an angle close to
$90^\circ$ to the SN explosion axis, no GRB is seen; one
observes {\it only} an XRF (X-ray Flash) and then a powerful UV flash 
caused by interaction in the shock and the envelope surrounding the pre-SN 
as was in the case of SN\,1993J.

    Thus, if an SN is observed close to the explosion equator (and this
situation is the most probable) and if there is a sufficiently dense 
stellar-wind envelope around a massive collapsing star core, then only
the shock breakout effect is to be observed in X-rays and in optical.
    In that case the contribution of GRB afterglow into a light curve
of a ``usual'' SN can be unnoticeable.
    One way or another, but it must be much less than for classical GRBs
observed close to the SN explosion axis (the least probable situation).
    In this connection, Filippenko et al. (2006)
noted recently that a substantially asymmetric explosion can be a genetic 
feature of core-collapse SNe of {\it all} types, though it is not clear yet 
if the mechanism generating the GRB is also responsible for the
star explosion.

   \section{Conclusion}

   The paper gives additional arguments in favor of the stellar-wind origin
of the shock breakout effect detected previously from Swift/XRT/UVOT data
(Campana et al., 2006) on XRF/GRB\,060218.
   In our optical spectra of the XRF/GRB\,060218 afterglow
we detected features interpreted as hydrogen lines which were also observed
in early ESO Lick, ESO VLT and NOT spectra
(Pian et al. 2006, Sollerman et al. 2006).
   Hydrogen was detected
in spectra of a GRB afterglow for the first time, which is a direct
sign of a relic wind envelope around a core-collapse progenitor star.

    Results of modelling two BTA spectra obtained in 2.55 and 3.55 days after
explosion of SN\,2006aj related to the X-ray flash XRF/GRB\,060218 are 
represented.
  The spectra were modeled in the Sobolev approximation with the
help of the SYNOW code (Branch et al. 2001; Elmhamdi et al. 2006).
In these early spectra of the Type Ic SN\,2006aj we detected spectral
features interpreted as:

 (1) the $H\alpha$ PCyg profile for velocities of 
$\sim$33000\,km\,s$^{-1}$ ---\ a wide and almost unnoticeable deformation 
of continuum in the range of $\approx5600 - 6600\AA$ for rest wavelengths 
(z=0) at the first epoch (2.55 days), and 

 (2) a part of the $H\alpha$ PCyg profile in absorption blueshifted by
24000\,km\,s$^{-1}$ ---\ a wide and low-contrast spectral feature at 
$\approx6100\AA$ (rest wavelength) at the second epoch (3.55 days).

    Evolution of the same spectral features can be traces also by spectra of
SN\,2006aj obtained with other telescopes
(Pian et al. 2006, Sollerman et al. 2006).
  Such $H\alpha$ lines can directly confirm the existence of a
stellar-wind envelope which was {\it already} observed during 
XRF/GRB\,060218 itself as a powerful black-body component first in X-rays 
and then in the optical spectrum --- the so-called  shock breakout effect 
(Campana et al., 2006).
  Thus, taking into account early observations
carried out with BTA and other telescopes (Table 1), it may be said that we 
observe evolution of optical spectra of the core-collapse 
SN\,2006aj --- {\it a transition} from the phase of the Colgate shock 
breakout effect to spectra of the phase of brightness increase 
corresponding to the radioactive heating.

    Our identification of these features with the $H\alpha$ hydrogen line in
early spectra of the Type Ic SN\,2006aj was also confirmed by spectral
observations of usual (non-identified with GRBs) Type Ic and Ib
core-collapse SNe and interpretation of their spectra (in which hydrogen 
was also found) with the SYNOW code (Branch et al. 2002; Branch et al. 2006; 
Elmhamdi et al. 2006).

    If interpretation of the thermal component in the spectrum of
GRB/XRF\,060218 as interaction between the SN shock and the wind envelope 
around the SN\,2006aj/XRF\,060218 progenitor star will be confirmed by 
observations of afterglow of other bursts, then it will give a new impulse 
to development of the theory of GRBs themselves and of the
core-collapse SNe.
  The intermediate redshift GRB/SNe are observed
relatively rarely (Chapman et al. 2007), but they are the most informative 
events (such as XRF\,/GRB\,060218/SN\,2006aj or GRB\,030329/SN\,2003dh) 
from the point of view of comprehension of relation between GRBs and SNe.
  And the key moment of the study of these transient sources
may be the search for manifestations of wind envelopes around core-collapse 
progenitor stars of GRBs both in early spectra and in the photometry of GRB 
afterglows.

\begin{acknowledgements}
 
   The authors are grateful to David Branch and other authors of the SYNOW
code for consultations on installation and practical usage of the code in 
SAO RAS and to E.L.Chentsov and S.N.Fabrika for reading and constructive 
critics of the first version of the text.
  The work was partially supported
by the Spanish Research Programs ESP2005-07714-C03-03 and AYA2004-01515.

\end{acknowledgements}

\begin{table*}[!p]
\setcaptionmargin{0mm} \onelinecaptionstrue
\captionstyle{flushleft}
\caption{
Early spectra of the supernova XRF\,060218/SN\,2006aj
obtained before Feb. 23 2006 UT with different telescopes.
The spectrum time is time after XRF/GRB\,060218 \textit{Swift} trigger:
T$_{sp}$ is in days after 2006 Feb. 18.149 UT.
Only spectra with a high signal/noise ratio are given.
Early spectra obtained by Modjaz et al. (2006) with the 1.5-meter telescope
FLWO in 3.97 days after the burst were not included because of their low
signal/noise ratio. }
\label{meanbe}
\bigskip
\begin{tabular}{c c c c l l l }     
\hline\hline
Telescope & T$_{sp}$\,and 2006 UT  & References \\
\hline
   MDM (2.4m) & 1.95 days (Feb.20.097) & Mirabal et al.2006 \\
   BTA (6m) & 2.55 days (Feb.20.70) & Fatkhullin et al. 2006 \\
   ESO VLT (8m) & 2.89 days (Feb.21.041) & Pian et al. 2006\\
   BTA (6m) & 3.55 days (Feb.21.70) & Fatkhullin et al. 2006 \\
   NOT (2.56m) & 3.78 days (Feb.21.93) & Sollerman et al. 2006\\
   ESO Lick (3m) & 4.01 days (Feb. 22.159) & Pian et al. 2006\\
   ESO VLT (8m)& 4.876 days (Feb. 23.026) & Pian et al. 2006\\
\hline
\end{tabular}
\end{table*}
%
%

\begin{table*}[!p]
\setcaptionmargin{0mm} \onelinecaptionstrue
\captionstyle{flushleft}
\caption{
  The table of model computation parameters in the case of layers undetached
from the photosphere for all elements and their ions 
(``the undetached case'' in Fig.\,\ref{spektr}).
  The parameters correspond to the synthetic spectrum represented
by the thick black line in Fig.\,\ref{undetached} (see the text).
  More detailed information about setting parameters for the SYNOW code
see in http://www.nhn.ou.edu/$\sim$parrent/synow.html
}
\label{catalog}
\bigskip
\begin{tabular}{lccccc}
\hline \hline

Parameters  & ``the undetached case'' \\

\hline
V$_{phot}$ \footnote{V$_{phot}$ is velocity of the photosphere in km/s. }
		       &33000\\
V$_{max}$\footnote{V$_{max}$ is the upper limit of velocities in the 
model. }
		       &70000\\
t$_{bb}$ \footnote{t$_{bb}$ is temperature of the photosphere in Kelvin 
degrees.}
		    &9000\\
ai      \footnote{ai are ionization stages of all ions considered in the 
model. }
&H I, He I, Fe III, Fe II, Si II, Si I, O I, Ca II, Ti II, CII, CI, NI, 
MgII, MgI, MnII, NeI\\
tau1      \footnote{tau1 are corresponding optical depths of line formation 
of every ion.}
&.2, .08, .04, .04, .0002, .0002, .002, .5, .02, .0005, .0005, .1, .0002, .
0002, .0002, .0005\\
vmine      \footnote{vmine are the least velocities {\it over the 
photosphere } for every ion 
(in 1000 km s$^{-1}$)}
      & 33.00 for all ions \\
ve      \footnote{ve are the characteristic velocityes ($v_e$)
in the used law $\tau(r) \sim exp(-v(r)/v_e)$)
of relation between optical depth of lines of a given ion $\tau$ and $v$,
where $v\sim r$.
}
      & 20.00 for all ions \\
\hline
\end{tabular}
\end{table*}

\begin{table*}[!p]
\setcaptionmargin{0mm} \onelinecaptionstrue
\captionstyle{flushleft}
\caption{
  Parameters of the SYNOW model for the case of layers of some elements
undetached from the photosphere in the expanding SN envelope 
(``the detached case'' in Fig.\,\ref{spektr}).
  The parameters correspond to one of the synthetic spectra shown in
Fig.\,\ref{figure1c} by the thick black line (see the text).
}
\label{ catalog }
\bigskip
\begin{tabular}{lccccc }     
\hline\hline
    
Parameters  & ``the detached case'' \\

\hline                    
V$_{phot}$     & 18000  \\  
V$_{max}$     &75000  \\
t$_{bb}$   & 8200    \\
ai
&HI,\,\, HeI,\,\, FeII,\,\, SiII,\,\, OI,\,\, CaI,\,\, CaII,\,\, TiII,\,\, 
CII,\,\, CI,\,\, NI,\,\, MgII,\,\, MgI,\,\, NaI\\
tau1 & 0.16,\,\, 0.02,\,\, 0.08, 0.03, 0.1, 0.3, 0.3, 0.02, 0.0004, 0.04, 
0.100, 0.005, 0.03, 0.02\\
vmine &24.00, 24.00, 24.00, 24.00, 18.00, 18.00, 18.0, 18.00, 24.0, 24.0, 
18.0, 24.0, 24.0, 18.00\\
ve&10.00, 20.00, 20.00, 20.00, 20.00, 10.00, 10.0, 20.00, 10.0, 10.0, 20.0, 
20.0, 20.0, 20.00\\
\hline
\end{tabular}
\end{table*}

\end{document}